\documentclass[conference,a4paper]{APSIPA2021}
\usepackage{multirow}
\usepackage[dvips]{graphicx}  
\usepackage{amsmath}
\usepackage{amsxtra}
\usepackage{threeparttable}

\def\BibTeX{{\rm B\kern-.05em{\sc i\kern-.025em b}\kern-.08em T\kern-.1667em\lower.7ex\hbox{E}\kern-.125emX}}

\begin{document}

\title{Effective data screening technique for \\crowdsourced speech intelligibility experiments: \\
Evaluation with IRM-based speech enhancement}

\author{%
\authorblockN{%
Ayako Yamamoto\authorrefmark{1},
Toshio Irino\authorrefmark{1},
Shoko Araki\authorrefmark{2},
Kenichi Arai\authorrefmark{2},\\
Atsunori Ogawa\authorrefmark{2},
Keisuke Kinoshita\authorrefmark{2}, and
Tomohiro Nakatani\authorrefmark{2}
}
\authorblockA{%
\authorrefmark{1}
Graduate School of Systems Engineering, Wakayama University, Japan \\
E-mail: yamamoto.ayako@g.wakayama-u.jp, irino@wakayama-u.ac.jp,  Tel: +81-73-457-8475}

\authorblockA{%
\authorrefmark{2}
NTT Corporation, Japan\\
E-mail: \{shoko.araki.pu,kenichi.arai.yw,atsunori.ogawa.gx,keisuke.kinoshita.mb,tomohiro.nakatani.nu\}@hco.ntt.co.jp  }
}


\maketitle
\thispagestyle{empty}

\begin{abstract}

It is essential to perform speech intelligibility (SI) experiments with human listeners in order to evaluate objective intelligibility measures for developing effective speech enhancement and noise reduction algorithms.
Recently, crowdsourced remote testing has become a popular means for collecting a massive amount and variety of data at a relatively small cost and in a short time. However, careful data screening is essential for attaining reliable SI data. 
We performed SI experiments on speech enhanced by an ``oracle'' ideal ratio mask (IRM) in a well-controlled laboratory and in crowdsourced remote environments that could not be controlled directly. We introduced simple tone pip tests, in which participants were asked to report the number of audible tone pips, to estimate their listening levels above audible thresholds. 
The tone pip tests were very effective for data screening to reduce the variability of crowdsourced remote results so that the laboratory results would become similar.
The results also demonstrated the SI of an oracle IRM, giving us the upper limit of the mask-based single-channel speech enhancement. 

\end{abstract}

\color{black}

\noindent\textbf{Index Terms}: Speech intelligibility, crowdsourced remote testing, listening level, speech enhancement, MVDR beamformer
\section{Introduction}
\label{sec:Introduction}


It is essential to perform speech intelligibility (SI) experiments with human listeners to evaluate objective intelligibility measures \,\cite{falk2015objective} in order to develop effective speech enhancement and noise reduction algorithms\,\cite{Loizou2013}.
Subjective SI experiments are usually performed in a soundproof room with well-controlled equipment in a laboratory.
Recently, crowdsourced remote testing has become a popular means of collecting a massive amount and variety of data at a relatively small cost and in a short time.
The novel coronavirus outbreak has promoted such crowdsourced experiments.
There have been several SI experiments using crowdsourcing services \cite{cooke2011crowdsourcing, cooke2013crowdsourcing,paglialonga2020automated,padilla2021binaural, yamamoto2021comparison,irino2022speech}, which is advantageous for gathering data from different age groups.
However, such reports are much less common than those on speech quality assessments \cite{buchholz2011crowdsourcing, ribeiro2011crowdmos,naderi2015effect, jimenez2018influence, naderi2020open}. 
This may be because it is almost impossible to control sound pressure level (SPL), acoustic environments including ambient noise, and listeners' hearing level, each of which affects the SI greatly.
To overcome the disadvantages of SI experiments, it is essential to establish effective methods to estimate all listeners' conditions for data screening.
The main purpose of this study is to compare laboratory and crowdsourced remote experiments on SI to establish an effective data screening technique.


\subsection{Evaluation target speech}
\label{sec:EvalTargetSpeech}


Selection of speech materials for subjective listening experiments is an important matter.
We used speech sounds processed by enhancement techniques that are essential for communication assistive devices for both normal hearing (e.g., \cite{DNSchallenge2022}) and hearing impaired listeners (e.g., \cite{clarity2021}).
In this paper, we evaluated the SI of noisy speech sounds enhanced with two speech enhancement techniques,
1) 
a single-channel ideal ratio mask (IRM) \cite{wang2014training} and 2) multi-channel mask-based beamformers \cite{Ito2017ICASSP} estimated by the IRM. 
An IRM is an ``oracle'' mask, which is defined by given clean speech and noise signals. An IRM itself, or enhanced speech with an oracle IRM, is widely used as a training target of deep neural networks (DNNs) for speech enhancement (e.g., \cite{wang2014training,MERL2014,PIT2017}).
We evaluated the original IRM \cite{wang2014training} to ascertain the baseline of the performance, although some phase-aware oracle masks have also recently been proposed\,(e.g.,~\cite{Xia2017UsingOR,cIRM2016}).
Multi-channel mask-based beamformers are realized by linear filters, and they can therefore enhance speech without any nonlinear distortion. Due to these distortionless characteristics, enhanced signals with a mask-based beamformer sound natural to the human ear and are also appropriate for preprocessing automatic speech recognition systems (e.g., \cite{Yoshioka2015ASRU,Heymann2016ICASSP,Erdogan2016IS}).

IRM is an oracle mask for single-channel speech enhancement, as described above, and is considered to provide us the upper performance limit for single-channel speech enhancement. In addition, a mask-based beamformer using an oracle IRM is considered to give a performance upper bound for a mask-based beamformer. However, there have been few reports on the subjective SI of these methods.  
We wanted to assess SI when using these two speech enhancement methods.

The rest of this paper is organized as follows. Section\,\ref{sec:SpeechMaterial} describes our speech material, including the enhanced speech signals. Section\,\ref{sec:Experiment} explains our testing method for conducting the crowdsourced remote testing and addresses the proposed data screening technique.
Section\,\ref{sec:ExpResult} then reveals our experimental results, and Section~\ref{sec:Conclusion} concludes this paper. 

 \color{black}

\section{Speech material with IRM}
\label{sec:SpeechMaterial}

This section explains speech materials used in the experiments. Source speech sounds, recorded in a real environment, were processed by speech enhancement algorithms with IRM. 

\subsection{Ideal Ratio Mask (IRM)}
\label{sec:IRM}


As a single-channel speech enhancement, this paper considers an oracle IRM based approach \cite{wang2014training} for investigating the upper limit of performance of single-channel speech enhancement. 
Let $x_i(n)$ be an observed noisy speech at  $i$-th microphone
\setlength{\abovedisplayskip}{2pt} 
\setlength{\belowdisplayskip}{2pt} 
\begin{equation}
 x_i(n) = s_i(n) + v_i(n) = h_{ij}(n)*c(n) + v_i(n), 
 \label{eq:xi=si+vi}
\end{equation}
where $n$ is a time index, and $c(n)$, $h_{ij}(n)$, and $v_i(n)$ are a clean speech, an impulse response from the sound source at $j$-th position to microphone $i$, and noise at microphone $i$, respectively. 
Its short time Fourier transform (STFT) is 
\begin{equation}
 x_{itf}=s_{itf}+v_{itf},  
\end{equation}
where $t$ and $f$ are time and frequency indices, respectively. 
The IRM for enhancing the speech signals is given as \cite{wang2014training}
\begin{equation}
 M_{tf}=\left(\frac{|s_{1tf}|^2}{|s_{1tf}|^2+|v_{1tf}|^2}\right)^{0.5},
 \label{eq:IRM}
\end{equation}
and the enhanced signal in the STFT domain with the IRM is 
 \begin{equation}
     y_{tf} = M_{tf}x_{1tf}.
 \end{equation}
The enhanced speech signal is obtained by applying the inverse STFT and the overlap-add to $y_{tf}$. 
We refer to this method as ``$\mathrm{MASK_{1ch}^{IRM}}$'' in this paper. 

\subsection{MVDR beamformer}
\label{sec:MVDR}

As a multi-channel speech enhancement, this paper investigates
a mask-based beamformer, recognized as the state-of-the-art front-end for recent
ASR systems 
\cite{Yoshioka2015ASRU,Heymann2016ICASSP,Erdogan2016IS}.

\color{red}

\color{black}

Let $\mathbf{x}_{tf}=[x_{1tf},\cdots,x_{Itf}]^{\mathsf{T} }$ 
be the observation vector with multiple microphones $i~\{i| 1 \le i \le I\}$ ($I=2$ in this paper) in the STFT domain, 
where $(\cdot)^{\mathsf{T} }$ denotes the transpose. 
The MVDR beamformer coefficients $\mathbf{w}_{f}=[w_{1f},\cdots,w_{If}]^{\mathsf{T}}$ are given as 
 \begin{equation}
 \mathbf{w}_f = \frac{{a_{rf}^*\mathbf{R}_{\mathbf{v}_f}^{-1}}\mathbf{a}_f}{\mathbf{a}_f^H{\mathbf{R}_{\mathbf{v}_f}^{-1}}\mathbf{a}_f}, 
  \label{eq:mvdr}
\end{equation}
where $\mathbf{a}_f$ is the (estimated) steering vector, $r$ is the reference microphone index, and 
$\mathbf{R}_{\mathbf{v}_f}=\frac{1}{T}\sum_t^T(1-M_{tf})\mathbf{x}_{tf}\mathbf{x}_{tf}^H$ (where $T$ is a number of frames) 
is the spatial covariance matrix (SCM) of noise. In addition, $(\cdot)^H$ and $(\cdot)^*$ denote the Hermitian transpose and complex conjugate, respectively.
The steering vector $\mathbf{a}_f$ in Eq. \ref{eq:mvdr} is given using the estimated SCMs for speech 
$\mathbf{R}_{\mathbf{s}_f}=\frac{1}{T}\sum_t^TM_{tf}\mathbf{x}_{tf}\mathbf{x}_{tf}^H$ 
and noise $\mathbf{R}_{\mathbf{v}_f}$ as \cite{Ito2017ICASSP}
\begin{equation}
 \mathbf{a}_f = \mathbf{R}_{\mathbf{v}_f}\mbox{maxeig}\left(\mathbf{R}_{\mathbf{v}_f}^{-1}\mathbf{R}_{\mathbf{s}_f}\right), 
 \end{equation}
where maxeig$(\mathbf{A})$ is the operation for finding the eigenvector corresponding to the largest eigenvalue of matrix $\mathbf{A}$. 

The enhanced signal in the STFT-domain with the MVDR beamformer is 
\begin{equation}
 y_{tf} = \mathbf{w}_{f}^H \mathbf{x}_{tf}. 
\end{equation}

To calculate the SCMs for the mask-based MVDR beamformer, we used the following two types of mask
\begin{itemize}
 \item ``$\mathrm{MVDR_{2ch}^{IRM}}$'': $M_{tf}$ is the IRM in Eq. \ref{eq:IRM}
 \item ``$\mathrm{MVDR_{2ch}^{EST}}$'': $M_{tf}$ is determined by the preset noise period $P_v$ (in this paper 288 msec.~at both the beginning and end of the noisy speech); i.e., 
 \[M_{tf}=
 \left\{
\begin{array}{ll}
0 & \mbox{for}~ t \in P_v \\
1 & \mbox{(otherwise)}
\end{array}
\right.\]
 \end{itemize}

\subsection{Recording babble noise and impulse response}
\label{sec:RecordIRBabble}

We used recorded babble noise for $v_i(n)$ in Eq. \ref{eq:xi=si+vi}. 
Mixed or single-speaker speech was played simultaneously from many loudspeakers on a large office floor, and they were recorded using two microphones with 4\,cm spacing in a conference room adjacent to it (reverberation time: approx. 360 ms; door open).
We also measured the impulse responses $h_{ij}(n)$ in Eq.~ \ref{eq:xi=si+vi} to the two microphones from 12 loudspeaker positions (index $j$) including nine positions with three directions ($-30^\circ$, $0^\circ$, $+30^\circ$ ) $\times$ three distances (70\,cm, 100\,cm, 130\,cm), and three locations with a distance of 90 cm and angles of $90^\circ$, $-75^\circ$, and $-105^\circ$.

\subsection{Clean and noisy speech signals}
\label{sec:SourceSpeech}



Clean speech $c(n)$  in Eq. \ref{eq:xi=si+vi} used for the subjective listening experiments were Japanese 4-mora words. They were uttered by a male speaker (label ID: mis) and drawn from a database of familiarity-controlled word lists, FW07 \cite{Kondo2007NTTTohoku}, which has been used in previous experiments \cite{yamamoto2020gedi, yamamoto2021comparison}.
The dataset contains 400 words for each of four familiarity ranks, and the average duration of a 4-mora word is approximately 700 \,ms. The source speech sounds were obtained from the word set with the least familiarity to prevent increment to the SI score by guessing commonly used words.

The observed noisy speech $x_i(n)$ was produced by convolving the clean speech $c(n)$ and an impulse response $h_{ij}$ of a source position $j$ and adding babble noise $v_i(n)$ as shown in Eq. \ref{eq:xi=si+vi}. 
The signal-to-noise ratio (SNR) conditions ranged from $-9$ to $+3$\ dB in $3$-dB steps.
These noisy speech signals were referred to as ``unprocessed''.
Three types of speech enhancement algorithms, $\mathrm{MASK_{1ch}^{IRM}}$, $\mathrm{MVDR_{2ch}^{IRM}}$, and $\mathrm{MVDR_{2ch}^{EST}}$, described in Secs.~\ref{sec:IRM} and ~\ref{sec:MVDR}, were applied to the unprocessed sounds.  The processing was performed at 16\,kHz sampling rate, and the sounds were then upsampled to 48\,kHz for playing on a web browser. 


\section{Experiment}
\label{sec:Experiment}

We used a set of web pages developed for previous experiments \cite{yamamoto2021comparison}; in addition, we introduced good practices to improve the quality of the experiments \cite{irino2022speech}.
One of these was to evaluate the listening condition by using tone pip tests as described in Section \ref{sec:TonePip}. Another one is to measure vocabulary size which may influence SI score.

\subsection{Experimental procedure}
\label{sec:ExpProcedure}

The main experiment was performed after a practice session to familiarize the participants with the experimental procedure.
The participants were instructed to write down the words that they heard using hiragana characters during a 4-second silent period until the presentation of the next word. The answers were filled in the provided answer sheets (PDF), which had been printed in advance. 
This was done on paper because it would be difficult to type the four mora words (consisting of approximately 6~-~9 Roman characters each) precisely into a computer within 4 seconds. 
After listening to ten words in one session, the participants were required to type the handwritten words into the answer columns on a web page (no time limit). The web page would count the number of hiragana characters and prompt the participants to fill in the answers correctly.

The total number of presented stimuli was 400 words, comprising a combination of the four enhancement conditions, \{unprocessed, $\mathrm{MASK_{1ch}^{IRM}}$, $\mathrm{MVDR_{2ch}^{IRM}}$, $\mathrm{MVDR_{2ch}^{EST}}$\} and the five SNR conditions with 20 words per condition. Each subject listened to a different word set, which was assigned randomly to avoid bias caused by word difficulty. 
The experiments were divided into two one-hour tasks to fulfill the crowdsourcing requirement of task duration.

\subsection{Laboratory experiments}
\label{sec:ExpLab}

In the laboratory experiments, the sounds were presented diotically via a DA converter (SONY, NW-A55) over headphones (SONY, MDR-1AM2). 
The SPL of the stimulus was 63\,dB in ${L_{Aeq}}$, which was calibrated with an artificial ear (Br\"{u}el \& Kj\ae r, Type\,4153) and a sound level meter (Br\"{u}el \& Kj\ae r, Type\,2250-L).
Listeners were seated in a sound-attenuated room (YAMAHA, AVITECS) with a background noise level of approximately 26dB in $L_{\rm Aeq}$.
Fourteen young Japanese normal hearing listeners (seven males and seven females, 20–24 years) participated in the experiments. 
The subjects were all naive to our SI experiments and had a hearing level of less than 20\,dB between 125\,Hz and 8000\,Hz.

\subsection{Remote experiments}
\label{sec:ExpRemote}

The experimental tasks were outsourced to a crowdsourcing service provided by Lancers Co. Ltd. in Japan \cite{Lancers}, as in \cite{yamamoto2021comparison}.
Any crowdworker could participate in the experimental task on a first-come-first-served basis. The participants were asked to perform the experiments in a quiet place and to set the volumes of their devices to an easily listenable level. The volume level was recorded and used to maintain consistency between the two experimental tasks, which a total of 39 participants completed.
There was a large variety of Japanese participants ages ( 21 -- 64 years) and there were three self-reported HI listeners. 

\subsection{Good practices for data screening}
\label{sec:GPdatascreen}

\begin{figure}[t]
\begin{center} 
    \centering
    
\includegraphics[width = 1\columnwidth] {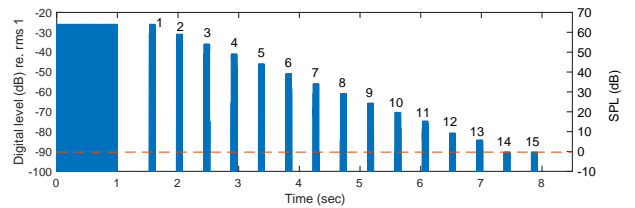}
         \vspace*{-8pt}
    \caption{RMS digital level of a sequence of 15 tone pips that decreases at 5\,dB steps. The right y-axis shows SPL when the first tone pip is assumed to be 64\,dB SPL.}
    \label{fig:TonePip}
   \vspace*{-10pt}

\end{center}
\end{figure}

\subsubsection{Tone pip test for estimating listening conditions}
\label{sec:TonePip}

As described in Section \ref{sec:Introduction}, it is essential to evaluate listening conditions for data screening to achieve reliable results. We introduced a web page to estimate how much the sound level was presented above the threshold, which was determined by the listener's absolute threshold, ambient noise level, and audio device.

A sequence of 15 tone pips with 5\,dB decreasing steps was presented to the listeners, who were asked to report the number of audible tone pips, $N_{pip}$.
Fig.\,\ref{fig:TonePip} shows the RMS digital level of the sequence of tone pips following a 1-second reference tone sound which has the same SPL, $L_{ref}$, as the stimulus speech sounds. The tone frequencies were 500\,Hz, 1000\,Hz, 2000\,Hz, and 4000\,Hz, to cover the speech range. 
Remote participants listened to the stimulus sounds in various environments, although they had been instructed to perform the task at a ``quiet'' place. $N_{pip}$ gives a rough estimate of the listening level $L_{lis}$ above the audible threshold at their individual acoustic environment as
\begin{eqnarray}
    L_{lis} & = & 5 \cdot(N_{pip}-1).
      \label{eq:Llis=5Npip}
\end{eqnarray}
$ L_{lis} $ is similar to, but not the same as, ``sensation level,'' which is defined as the relative level to the absolute threshold of the listener and has been used in many psychoacoustic experiments\,\cite{moore2013introduction}. 
The SPL of the tone pip at the threshold is calculated as
\begin{eqnarray}
    L_{pip}^{(th)} & = & L_{ref} - L_{lis}.
      \label{eq:Lpip=Lref-Llis}
\end{eqnarray}
\color{black}
The right y-axis in Fig.\,\ref{fig:TonePip} shows the SPL when $L_{ref}$ is 64\,dB SPL, as an example. Although we could not know the SPLs of sounds presented to participants in the remote experiments, it was possible to use the listening level, $L_{lis}$, for data screening effectively, as described in Section \ref{sec:DataScreening}. 
The tone pip test took only a few minutes and have been introduced in another SI experiment\,\cite{irino2022speech}.
The procedure is very simple and can be improved by using an ascending sequence together.

\begin{figure*}[t] 
\begin{minipage} {0.49\hsize}
    \centering
    \includegraphics[width = 0.9\columnwidth]
    {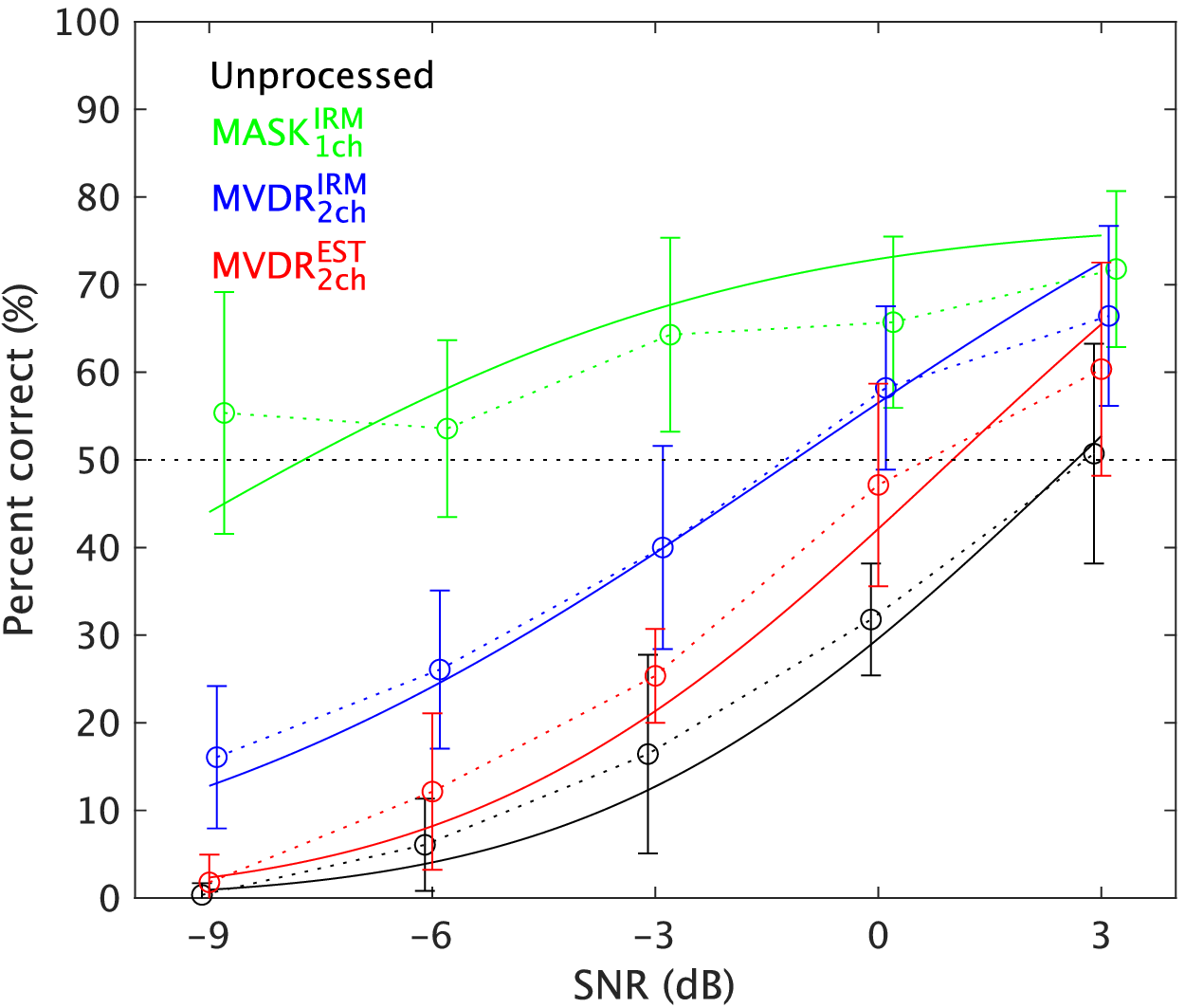}
\end{minipage}     
\begin{minipage} {0.49\hsize}    
    \includegraphics[width = 0.9\columnwidth]
    {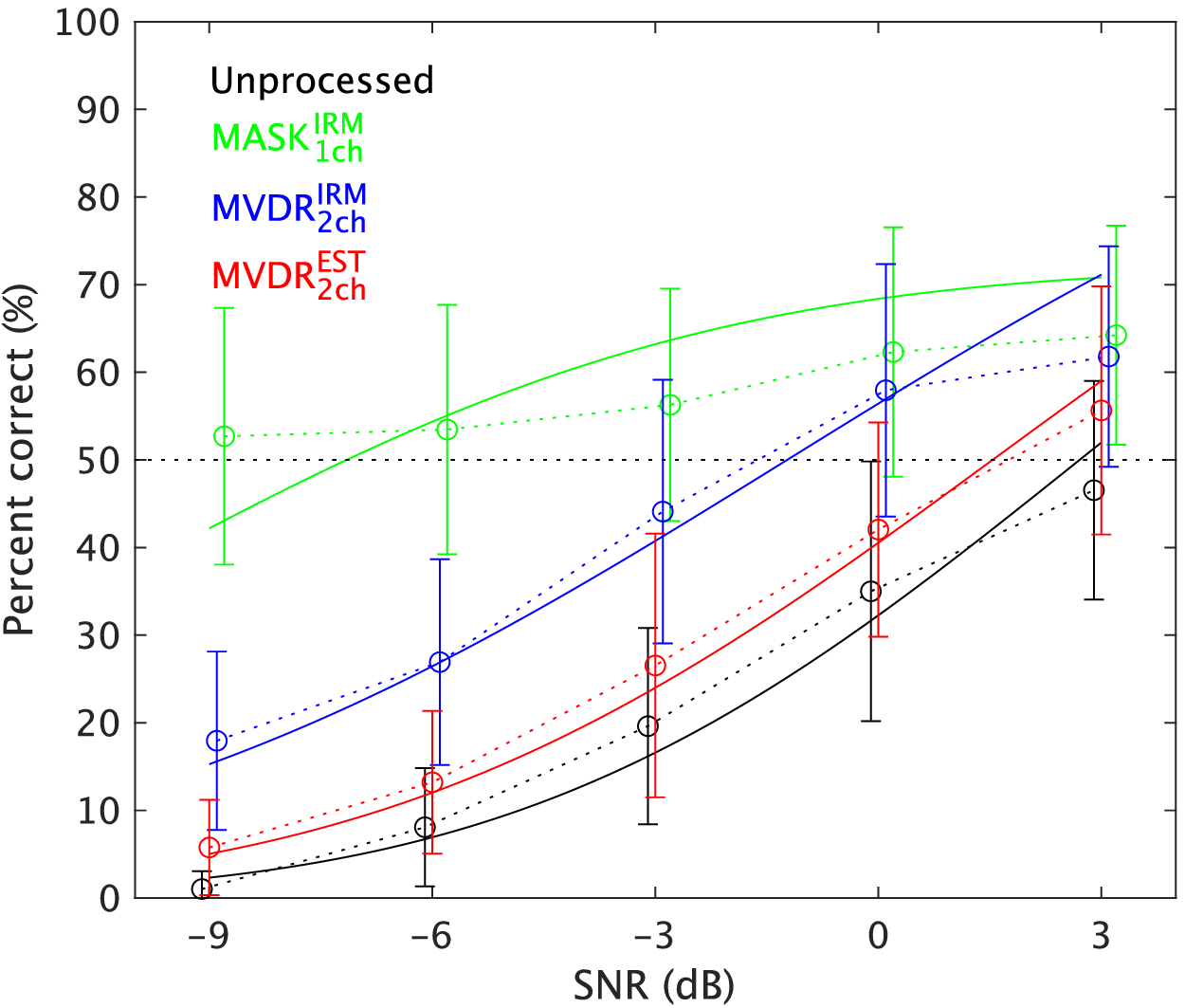} 
\end{minipage}  
\caption{Mean and standard deviation (SD) of word correct rates (\%) across participants in the laboratory (left) and crowdsourced remote experiments (right). Solid lines show the psychometric functions estimated with the cumulative Gaussian function.}
\label{fig:Exp_PsychoFunc}
\end{figure*}

\begin{figure*}[ht]
\begin{tabular}{ccc}
\begin{minipage} {0.33\hsize}
    \centering
    \includegraphics[width=1\linewidth]{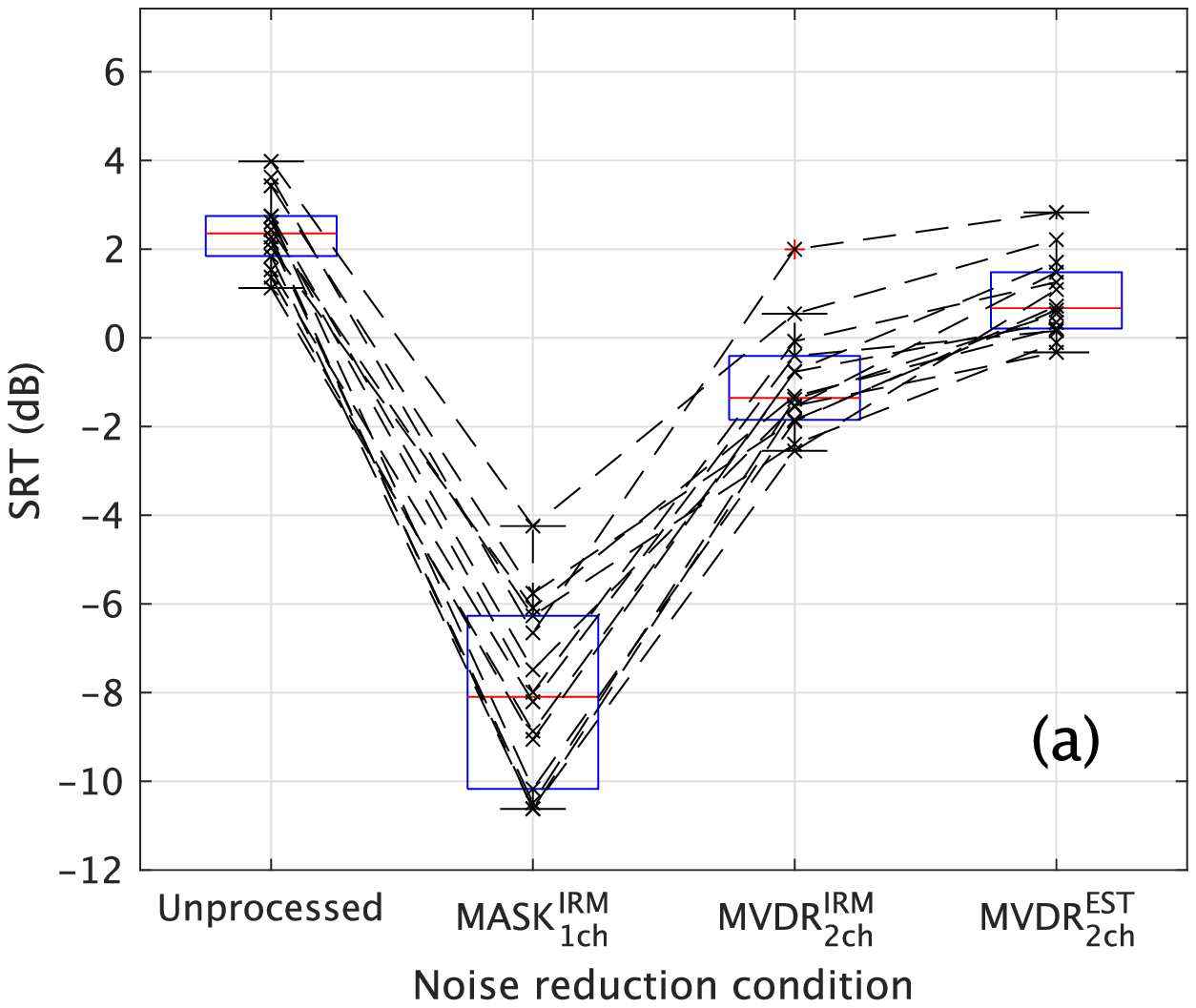}
    \label{fig:SRTBox_ExpLab}
\end{minipage}

\begin{minipage} {0.33\hsize}
    \centering
    \includegraphics[width=1\linewidth]{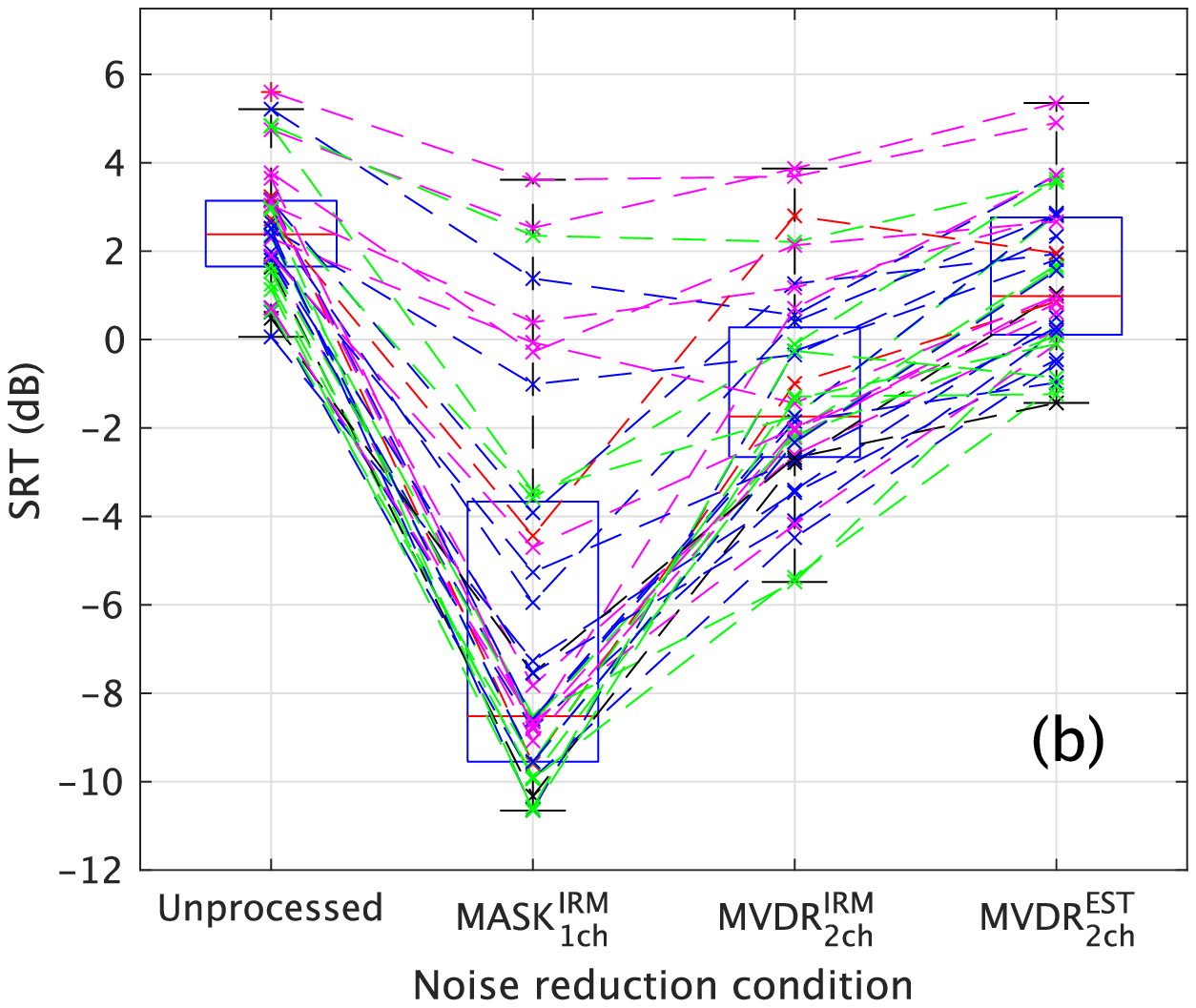}
    \label{fig:SRTBox_ExpRemote}
\end{minipage}

\begin{minipage} {0.33\hsize}
    \centering
    \includegraphics[width=1\linewidth]{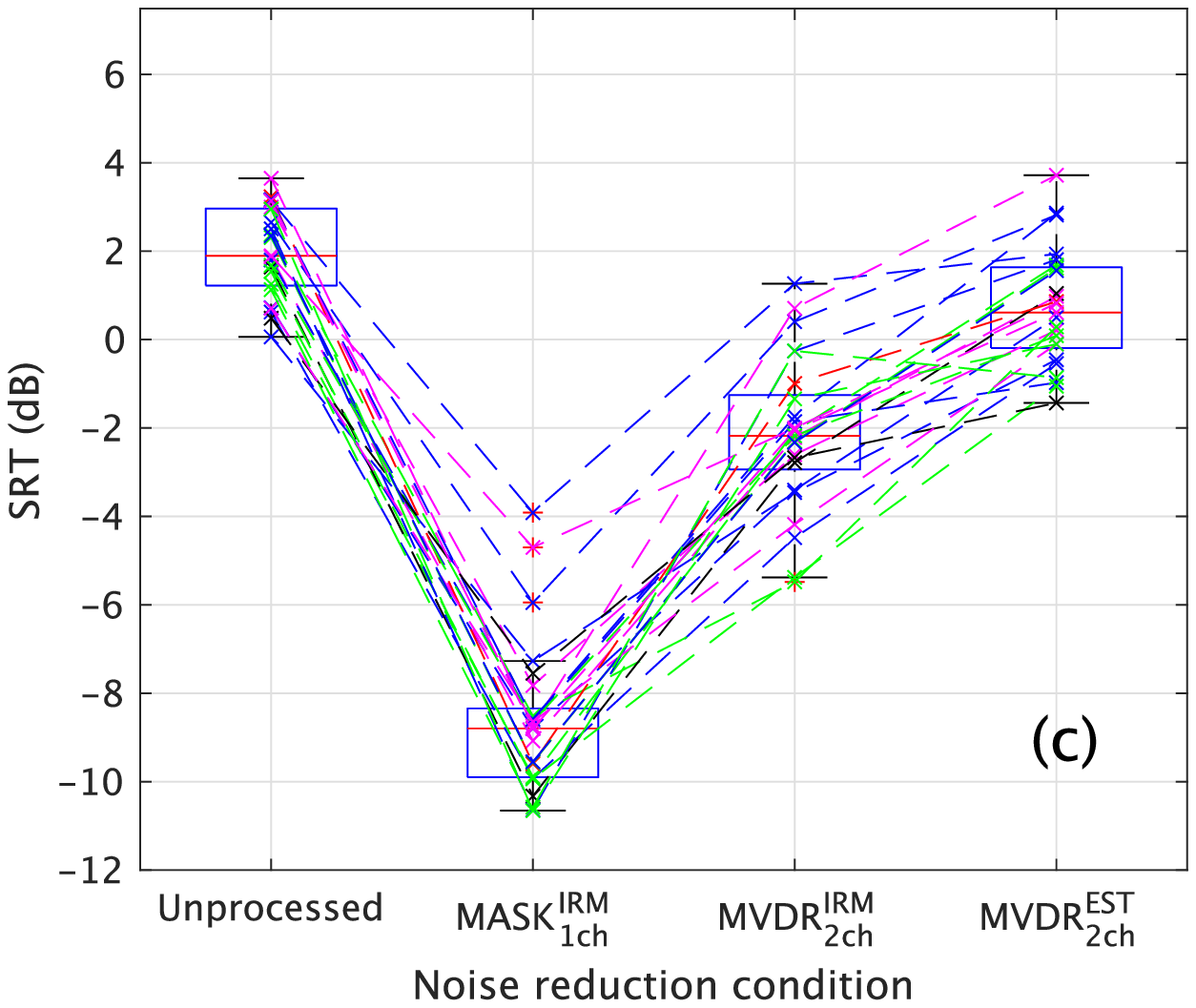}
    \label{fig:SRTBox_ExpRemoteDataScr}
\end{minipage}
\end{tabular}
\caption{SRT values in dB. (a) Laboratory experiments (N=14); (b) crowdsourced remote  experiments (N=39); (c) crowdsourced remote  experiments after data screening (N=25). Dashed lines show individual results. Color represents the age range: (black) 20s, (green) 30s, (blue) 40s, (magenta) 50s, (red) 60s.}
\label{fig:SRTBox}
\end{figure*} 

\subsubsection{Vocabulary size}
Another factor influencing the SI is listener's vocabulary size. 
We also introduced a test to estimate the vocabulary size of Japanese participants \cite{kondo2013hyakurakan} after the listening tasks. The collected data will be analyzed in the next study.

\section{Results}
\label{sec:ExpResult}

We first performed data cleansing of participants' answers to calculate the SI, as described in \cite{yamamoto2021comparison}, and then we compared the results of the laboratory and remote experiments.

\subsection{Psychometric function of speech intelligibility}
\label{sec:SpIntel}

Figure \ref{fig:Exp_PsychoFunc} shows the word correct rates of the laboratory and remote experiments as a function of the SNR. Circles and error bars represent the mean and standard deviation (SD) across participants. The solid curves represent psychometric functions estimated with the cumulative Gaussian function using psignifit, a Bayesian method \cite{schutt2016painfree}. 
The psychometric functions of both experiments were virtually the same. This implies that the remote experiment results could be usable instead of the reliable laboratory results. Further analysis will be provided later.
The psychometric functions of the enhanced conditions ($\mathrm{MASK_{1ch}^{IRM}}$, 
$\mathrm{MVDR_{2ch}^{IRM}}$, $\mathrm{MVDR_{2ch}^{EST}}$) were
located higher than those of the unprocessed condition, implying these algorithms were very effective for speech enhancement. In contrast, spectral subtraction and Wiener filter algorithms did not improve the SI more than the unprocessed condition as observed in the previous studies\,\cite{yamamoto2021comparison,yamamoto2020gedi}.
Furthermore, the psychometric function of $\mathrm{MASK_{1ch}^{IRM}}$ was greater than 50\% even in the worst SNR of $-9\,\mathrm{dB}$ and was not much affected by the SNR conditions. This implies that noise components were suppressed sufficiently to reduce masking effect.

It should be noted that Fig.~\ref{fig:Exp_PsychoFunc} gives the SI of speech sounds enhanced by the IRM and IRM-based beamformer for the first time with respect to a Japanese corpus.
The IRM is an ``oracle'' method that has been used in a training target of DNN enhancement methods\,\cite{wang2014training}. The results give important information about the upper limit or the goal of the subjective SIs derived by such enhancement methods.


\subsection{Speech reception threshold}
\label{sec:SRT}

Speech reception threshold (SRT) was calculated as the SNR value at which the psychometric function reaches a 50\% word correct rate. 
Figure \ref{fig:SRTBox}(a) shows SRTs across the noise reduction conditions of 14 participants in the laboratory experiments and Fig. \ref{fig:SRTBox}(b) shows those of 39 participants in the remote experiments. 
Each dashed line shows the individual result.
The tendency of the SRT variation across the conditions was similar.
However, the variability of the SRTs in the remote experiments was much greater than those in the laboratory experiments, particularly in the $\mathrm{MASK_{1ch}^{IRM}}$ condition. One of the reasons for this would be minimal controllability of listening conditions in the remote experiments. Another reason would be the different population of participants: those in the laboratory experiments were in their twenties alone, while there were participants ranging from their twenties to their sixties in the remote experiments. We need to specify the reason for effective data screening.





\begin{figure}[t]
\begin{center} 
    \centering
    \includegraphics[width = 0.85\columnwidth]
    {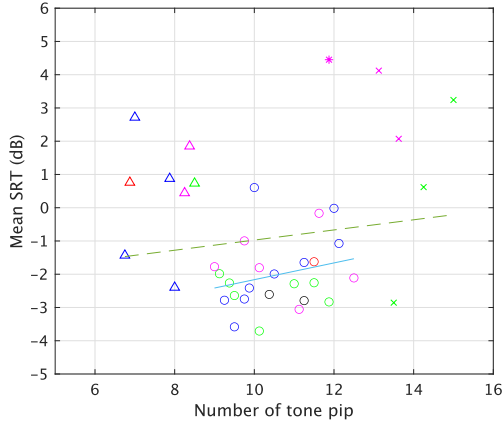}
    \caption{Relationship between the mean value of audible tone pips, $\bar N_{pip}$, and the mean SRT value (dB). $\triangle$ : $\bar N_{pip} < 9$; $\bigcirc$: $9\le \bar N_{pip} \le 13$; $\times$: $\bar N_{pip} > 13$; *: the outlier in Fig.\,\ref{fig:SRTBox}(b). There are three regression lines. Colors represent the age ranges in the same way as in Fig.\,\ref{fig:SRTBox}.}
     \label{fig:TonePipVsSRT_Remote}
\end{center}

\end{figure}

\subsection{Relationship between the number of tone pips and SRT}
\label{sec:Level_SRT}

As described in Section\,\ref{sec:TonePip}, we introduced tone pip listening tests.  
We surveyed the relationship between $N_{pip}$ and the SRT values for individual listeners.
Figure \ref{fig:TonePipVsSRT_Remote} shows a scatter plot between $N_{pip}$ averaged across four tone frequencies, $\bar N_{pip}$, and the SRT value averaged across the four enhancement conditions in the remote experiments.
There was very large variability between the participants. 
The green dashed line shows the regression line for all data points.
There was weak positive correlation between them, but it was not significantly different from no correlation ($r=0.14; p=0.39$). 
The high SRT values were observed only in the low and high ranges of $\bar N_{pip}$ but not in the middle range. 

When $\bar N_{pip}$ is very high, the participants' responses could be assumed to be suspicious. For example, when $\bar N_{pip} = 13$ and $L_{ref} = 64$ (dB) in Eq.\,\ref{eq:Lpip=Lref-Llis}, $L_{pip}^{(th)} = 4$ (dB), which is less than the average absolute threshold of normal hearing listeners as defined in the specification of audiometers \cite{ansi2018specification}, where the SPLs are 13.5\,dB, 7.5\,dB, 9.0\,dB, and 12.0\,dB at 500\,Hz, 1000\,Hz, 2000\,Hz, and 4000\,Hz, respectively. 
Moreover, the ambient noise level in a participant's ``quiet'' place may increase the threshold.  Consequently, we could assume the participants did not understand the experiment instructions or did not perform the listening tests honestly. 
On the other hand, when $N_{pip} < 9$, the listening level $L_{lis}$ in Eq.\,\ref{eq:Llis=5Npip} is less than 40\,dB, which may be insufficient for recognizing low-level consonants. 
We therefore assumed that the hearing thresholds and listening conditions, including ambient noise level, were different from the well-controlled one in the laboratory.

\subsection{Data screening}
\label{sec:DataScreening}
Based on the above discussion, we performed data screening by eliminating points with $\bar N_{pip} < 9$, $\bar N_{pip} > 13$, and the outlier in Fig.\,\ref{fig:SRTBox}(b).
There were 25 remaining data points, which are shown as circles in Fig.\,\ref{fig:TonePipVsSRT_Remote}. 
The blue solid line
shows the regression line, and there was no correlation ($r=0.25; p=0.22$).

\begin{figure}[t] 
\begin{center} 
    \centering
    \includegraphics[width =1\columnwidth] {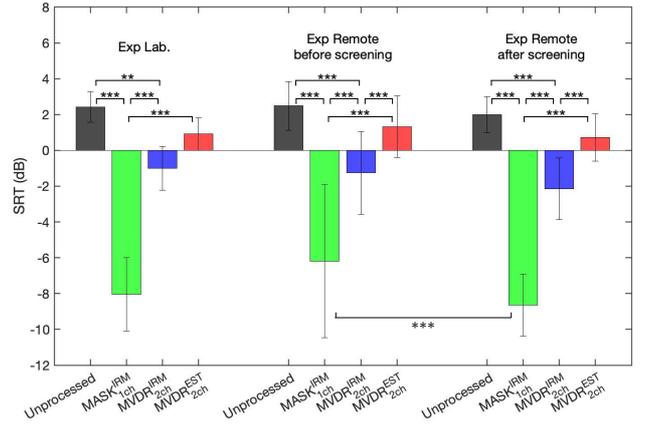}
    \caption{Mean and SD of the SRT (dB) across participants. Bars correspond to the four speech enhancement conditions. Experimental results in the laboratory (left, N=14), and in the crowdsourced remote experiments before the screening (middle, N=39 ) and after the screening (right, N=25). 
    We only labelled asterisks in the significant and meaningful combinations. 
    ***: $p < 0.001$; **: $p < 0.01$; *: $p < 0.05$.
    }
    \label{fig:SRTBar_ExpLabRemote}
\end{center} 
\end{figure}

\subsubsection{Result of data screening}

Figure \ref{fig:SRTBox}(c) shows that the SRT values across the enhancement condition for 25 participants remained after the above data screening. 
The variability of the SRT values was much lower than those of all 35 participants in Fig. \ref{fig:SRTBox}(b). The distributions across the enhancement conditions became similar to those in the laboratory experiments in Fig. \ref{fig:SRTBox}(a).
Overall, the procedure using $\bar N_{pip}$ was  effective for data screening to equalize the remote results to the laboratory results.


\subsubsection{Statistical result}


Figure\,\ref{fig:SRTBar_ExpLabRemote} shows the mean and SD of the SRT values across participants.
The left four bars show the experimental results of the four speech enhancement conditions in the laboratory. The middle four bars show the results before the data screening in the crowdsourced remote experiments. The right four bars show those after the data screening. As a result of multiple comparison tests, there were no significant differences between the same speech enhancement conditions except for $\mathrm{MASK_{1ch}^{IRM}}$  ($p<0.001$) in the remote experiments. 
The mean and SD values in the laboratory were similar to those after the screening.
In contrast, the SD of $\mathrm{MASK_{1ch}^{IRM}}$ before the screening was about twice as large as those in the laboratory and after the screening. This result demonstrates that the data screening successfully reduced variability across the participants. 

\color{black}

\section{Conclusion}
\label{sec:Conclusion}
In this study, we performed laboratory and crowdsourced remote experiments to evaluate the SI of noisy speech sounds enhanced by a single-channel IRM and multi-channel mask-based beamformers.
Although the SI scores were similarly improved by these enhancement methods, the variability was much greater in the remote experiment than in the laboratory experiment, probably because of the difference in the listening conditions. 
The tone pip test effectively reduced the variability to attain reliable data from crowdsourced remote experiments. It is an effective data screening technique that will be useful for the further studies.
Moreover, the SI scores of the IRM method implied the upper limit for single-channel speech enhancement because the IRM is an  ``oracle'' method.

\color{black}





\section{Acknowledgements}
This research was supported by JSPS KAKENHI Nos. 21H03468, and 21K19794.

  \bibliographystyle{IEEEtran}
  \bibliography{Reference_InterSp22.bib}

\begin{thebibliography}{10}
\providecommand{\url}[1]{#1}
\csname url@samestyle\endcsname
\providecommand{\newblock}{\relax}
\providecommand{\bibinfo}[2]{#2}
\providecommand{\BIBentrySTDinterwordspacing}{\spaceskip=0pt\relax}
\providecommand{\BIBentryALTinterwordstretchfactor}{4}
\providecommand{\BIBentryALTinterwordspacing}{\spaceskip=\fontdimen2\font plus
\BIBentryALTinterwordstretchfactor\fontdimen3\font minus
  \fontdimen4\font\relax}
\providecommand{\BIBforeignlanguage}[2]{{%
\expandafter\ifx\csname l@#1\endcsname\relax
\typeout{** WARNING: IEEEtran.bst: No hyphenation pattern has been}%
\typeout{** loaded for the language `#1'. Using the pattern for}%
\typeout{** the default language instead.}%
\else
\language=\csname l@#1\endcsname
\fi
#2}}
\providecommand{\BIBdecl}{\relax}
\BIBdecl

\bibitem{falk2015objective}
\BIBentryALTinterwordspacing
T.~H. Falk, V.~Parsa, J.~F. Santos, K.~Arehart, O.~Hazrati, R.~Huber, J.~M.
  Kates, and S.~Scollie, ``Objective quality and intelligibility prediction for
  users of assistive listening devices: Advantages and limitations of existing
  tools,'' \emph{IEEE signal processing magazine}, vol.~32, no.~2, pp.
  114--124, 2015. [Online]. Available:
  \url{https://ieeexplore.ieee.org/abstract/document/7038268/}
\BIBentrySTDinterwordspacing

\bibitem{Loizou2013}
P.~C. Loizou, \emph{{Speech Enhancement: Theory and Practice}}, 2nd~ed.\hskip
  1em plus 0.5em minus 0.4em\relax CRC Press, 2013.

\bibitem{cooke2011crowdsourcing}
M.~Cooke, J.~Barker, M.~L.~G. Lecumberri, and K.~Wasilewski, ``Crowdsourcing
  for word recognition in noise,'' in \emph{Proc. Interspeech 2011, Twelfth
  Annual Conference of the International Speech Communication Association},
  2011.

\bibitem{cooke2013crowdsourcing}
\BIBentryALTinterwordspacing
M.~Cooke, J.~Barker, M.~G. Lecumberri, and K.~Wasilewski, ``Crowdsourcing in
  speech perception,'' \emph{Crowdsourcing for speech processing: Applications
  to data collection, transcription and assessment}, vol. 137, p. 172, 2013.
  [Online]. Available: \url{https://doi.org/10.1002/9781118541241}
\BIBentrySTDinterwordspacing

\bibitem{paglialonga2020automated}
\BIBentryALTinterwordspacing
A.~Paglialonga, E.~M. Polo, M.~Zanet, G.~Rocco, T.~van Waterschoot, and
  R.~Barbieri, ``An automated speech-in-noise test for remote testing:
  Development and preliminary evaluation,'' \emph{American Journal of
  Audiology}, vol.~29, no.~3S, pp. 564--576, 2020. [Online]. Available:
  \url{https://doi.org/10.1044/2020\_AJA-19-00071}
\BIBentrySTDinterwordspacing

\bibitem{padilla2021binaural}
\BIBentryALTinterwordspacing
A.~Padilla-Ortiz and F.~Ordu{\~n}a-Bustamante, ``Binaural speech
  intelligibility tests conducted remotely over the internet compared with
  tests under controlled laboratory conditions,'' \emph{Applied Acoustics},
  vol. 172, p. 107574, 2021. [Online]. Available:
  \url{https://doi.org/10.1016/j.apacoust.2020.107574}
\BIBentrySTDinterwordspacing

\bibitem{yamamoto2021comparison}
\BIBentryALTinterwordspacing
A.~Yamamoto, T.~Irino, K.~Arai, S.~Araki, A.~Ogawa, K.~Kinoshita, and
  T.~Nakatani, ``Comparison of remote experiments using crowdsourcing and
  laboratory experiments on speech intelligibility,'' in \emph{Proc.
  Interspeech 2021}, 2021, pp. 181--185. [Online]. Available:
  \url{https://www.doi.org/10.21437/Interspeech.2021-174}
\BIBentrySTDinterwordspacing

\bibitem{irino2022speech}
\BIBentryALTinterwordspacing
T.~Irino, T.~Honoka, and A.~Yamamoto, ``Speech intelligibility of simulated
  hearing loss sounds and its prediction using the gammachirp envelope
  similarity index ({GESI}),'' in \emph{Proc. Interspeech 2022}, 2022.
  [Online]. Available: \url{https://arxiv.org/abs/2206.06573}
\BIBentrySTDinterwordspacing

\bibitem{buchholz2011crowdsourcing}
\BIBentryALTinterwordspacing
S.~Buchholz and J.~Latorre, ``Crowdsourcing preference tests, and how to detect
  cheating,'' in \emph{Proc. Interspeech 2011, Twelfth annual conference of the
  international speech communication association}, 2011. [Online]. Available:
  \url{https://www.isca-speech.org/archive/archive\_papers/interspeech\_2011/i11\_3053.pdf}
\BIBentrySTDinterwordspacing

\bibitem{ribeiro2011crowdmos}
\BIBentryALTinterwordspacing
F.~Ribeiro, D.~Flor{\^e}ncio, C.~Zhang, and M.~Seltzer, ``Crowdmos: An approach
  for crowdsourcing mean opinion score studies,'' in \emph{2011 IEEE
  international conference on acoustics, speech and signal processing
  (ICASSP)}.\hskip 1em plus 0.5em minus 0.4em\relax IEEE, 2011, pp. 2416--2419.
  [Online]. Available:
  \url{https://ieeexplore.ieee.org/abstract/document/5946971/}
\BIBentrySTDinterwordspacing

\bibitem{naderi2015effect}
\BIBentryALTinterwordspacing
B.~Naderi, T.~Polzehl, I.~Wechsung, F.~K{\"o}ster, and S.~M{\"o}ller, ``Effect
  of trapping questions on the reliability of speech quality judgments in a
  crowdsourcing paradigm,'' in \emph{Proc. Interspeech 2015, Sixteenth annual
  conference of the international speech communication association}, 2015.
  [Online]. Available:
  \url{https://www.isca-speech.org/archive/interspeech\_2015/i15\_2799.html}
\BIBentrySTDinterwordspacing

\bibitem{jimenez2018influence}
\BIBentryALTinterwordspacing
R.~Z. Jim{\'e}nez, L.~F. Gallardo, and S.~M{\"o}ller, ``Influence of number of
  stimuli for subjective speech quality assessment in crowdsourcing,'' in
  \emph{2018 Tenth international conference on quality of multimedia experience
  (QoMEX)}.\hskip 1em plus 0.5em minus 0.4em\relax IEEE, 2018, pp. 1--6.
  [Online]. Available:
  \url{https://ieeexplore.ieee.org/abstract/document/8463298}
\BIBentrySTDinterwordspacing

\bibitem{naderi2020open}
\BIBentryALTinterwordspacing
B.~Naderi and R.~Cutler, ``An open source implementation of itu-t
  recommendation p. 808 with validation,'' \emph{arXiv preprint
  arXiv:2005.08138}, 2020. [Online]. Available:
  \url{https://arxiv.org/abs/2005.08138}
\BIBentrySTDinterwordspacing

\bibitem{DNSchallenge2022}
H.~Dubey, V.~Gopal, R.~Cutler, A.~Aazami, S.~Matusevych, S.~Braun, S.~E.
  Eskimez, M.~Thakker, T.~Yoshioka, H.~Gamper, and R.~Aichner, ``{ICASSP} 2022
  deep noise suppression challenge,'' in \emph{ICASSP 2022 - 2022 IEEE
  International Conference on Acoustics, Speech and Signal Processing
  (ICASSP)}, 2022, pp. 9271--9275.

\bibitem{clarity2021}
S.~Graetzer, J.~Barker, T.~J. Cox, M.~Akeroyd, J.~F. Culling, G.~Naylor,
  E.~Porter, and R.~V. Munoz, ``Clarity-2021 challenges: {M}achine learning
  challenges for advancing hearing aid processing,'' in \emph{Proc. Interspeech
  2021}, 2021, pp. 686--690.

\bibitem{wang2014training}
Y.~Wang, A.~Narayanan, and D.~Wang, ``On training targets for supervised speech
  separation,'' \emph{IEEE/ACM Transactions on Audio, Speech, and Language
  Processing}, vol.~22, no.~12, pp. 1849--1858, 2014.

\bibitem{Ito2017ICASSP}
N.~Ito, S.~Araki, M.~Delcroix, and T.~Nakatani, ``{Probabilistic spatial
  dictionary based online adaptive beamforming for meeting recognition in noisy
  and reverberant environments},'' in \emph{IEEE International Conference on
  Acoustics, Speech and Signal Processing (ICASSP) 2017}.\hskip 1em plus 0.5em
  minus 0.4em\relax IEEE, 2017, pp. 681--685.

\bibitem{MERL2014}
F.~Weninger, J.~R. Hershey, J.~Le~Roux, and B.~Schuller, ``Discriminatively
  trained recurrent neural networks for single-channel speech separation,'' in
  \emph{2014 IEEE Global Conference on Signal and Information Processing
  (GlobalSIP)}, 2014, pp. 577--581.

\bibitem{PIT2017}
M.~Kolbaek, D.~Yu, Z.-H. Tan, J.~Jensen, M.~Kolbaek, D.~Yu, Z.-H. Tan, and
  J.~Jensen, ``Multitalker speech separation with utterance-level permutation
  invariant training of deep recurrent neural networks,'' \emph{IEEE/ACM Trans.
  Audio, Speech and Lang. Proc.}, vol.~25, no.~10, p. 1901–1913, oct 2017.

\bibitem{Xia2017UsingOR}
S.~Xia, H.~Li, and X.~Zhang, ``Using optimal ratio mask as training target for
  supervised speech separation,'' in \emph{APSIPA ASC2017}, 2017, pp. 163--166.

\bibitem{cIRM2016}
D.~S. Williamson, Y.~Wang, and D.~Wang, ``Complex ratio masking for monaural
  speech separation,'' \emph{IEEE/ACM Transactions on Audio, Speech, and
  Language Processing}, vol.~24, no.~3, pp. 483--492, 2016.

\bibitem{Yoshioka2015ASRU}
T.~Yoshioka \emph{et~al.}, ``{The NTT CHiME-3 system: Advances in speech
  enhancement and recognition for mobile multi-microphone devices},'' in
  \emph{Proceedings of Automatic Speech Recognition and Understanding
  (ASRU)}.\hskip 1em plus 0.5em minus 0.4em\relax IEEE, 2015, pp. 436--443.

\bibitem{Heymann2016ICASSP}
J.~Heymann, L.~Drude, and R.~Haeb-Umbach, ``{Neural network based spectral mask
  estimation for acoustic beamforming},'' in \emph{IEEE International
  Conference on Acoustics, Speech and Signal Processing (ICASSP) 2016}.\hskip
  1em plus 0.5em minus 0.4em\relax IEEE, 2016, pp. 196--200.

\bibitem{Erdogan2016IS}
H.~Erdogan, J.~R. Hershey, S.~Watanabe, M.~I. Mandel, and J.~L. Roux,
  ``{Improved {MVDR} beamforming using singlechannel mask prediction
  networks},'' in \emph{Proc. Interspeech 2017}.\hskip 1em plus 0.5em minus
  0.4em\relax IEEE, 2016, pp. 1981--1985.

\bibitem{Kondo2007NTTTohoku}
\BIBentryALTinterwordspacing
K.~Kondo, S.~Amano, Y.~Suzuki, and S.~Sakamoto, ``{NTT-Tohoku University
  Familiarity-Controlled Word Lists 2007 (FW07)},'' 2007. [Online]. Available:
  \url{http://research.nii.ac.jp/src/en/FW07.html}
\BIBentrySTDinterwordspacing

\bibitem{yamamoto2020gedi}
\BIBentryALTinterwordspacing
K.~Yamamoto, T.~Irino, S.~Araki, K.~Kinoshita, and T.~Nakatani, ``{GEDI}:
  Gammachirp envelope distortion index for predicting intelligibility of
  enhanced speech,'' \emph{Speech Communication}, vol. 123, pp. 43--58, 2020.
  [Online]. Available: \url{https://doi.org/10.1016/j.specom.2020.06.001}
\BIBentrySTDinterwordspacing

\bibitem{Lancers}
\BIBentryALTinterwordspacing
``Lancers co. ltd.'' [Online]. Available: \url{https://www.lancers.jp}
\BIBentrySTDinterwordspacing

\bibitem{moore2013introduction}
B.~C.~J. Moore, \emph{{An Introduction to the Psychology of Hearing}},
  6th~ed.\hskip 1em plus 0.5em minus 0.4em\relax Leiden, The Netherlands:
  Brill, 2013.

\bibitem{kondo2013hyakurakan}
T.~Kondo and S.~Amano, ``Hyakurakan: A test for chinese character to control
  the vocabulary variability of experimental participants ({in Japanese}),'' in
  \emph{Report of Japan Cognitive Sci. Soc.}, vol. JCSS-TR-69, 2013.

\bibitem{schutt2016painfree}
\BIBentryALTinterwordspacing
H.~H. Sch{\"u}tt, S.~Harmeling, J.~H. Macke, and F.~A. Wichmann, ``Painfree and
  accurate bayesian estimation of psychometric functions for (potentially)
  overdispersed data,'' \emph{Vision research}, vol. 122, pp. 105--123, 2016.
  [Online]. Available: \url{https://github.com/wichmann-lab/psignifit/wiki}
\BIBentrySTDinterwordspacing

\bibitem{ansi2018specification}
\BIBentryALTinterwordspacing
{\rm ANSI/ASA \;S3.\;6 - 2018}, ``Specification for audiometers,'' 2018.
  [Online]. Available:
  \url{https://webstore.ansi.org/standards/asa/ansiasas32018}
\BIBentrySTDinterwordspacing

\end{thebibliography}

\end{document}